\documentclass[10pt]{article}
\usepackage{latexsym}
\usepackage{graphicx}
\usepackage{multirow}
\usepackage{amsmath}
\usepackage{ulem}
\usepackage{setspace}
\usepackage{subfigure}
\begin{document}
\begin{center}\Large{A model of possible gravitational confinement of fast light particles}
\end{center}
\begin{center}
\renewcommand\thefootnote{\fnsymbol{footnote}}
Constantinos G. Vayenas \footnote{E-mail: cgvayenas@upatras.gr} \& Stamatios Souentie
\end{center}
\begin{center}
\textit{LCEP, Caratheodory 1, St., University of Patras, Patras GR 26500, Greece}
\end{center}
\begin{abstract}
According to special relativity and the equivalence principle, the Newtonian gravitational force between two particles with relativistic velocities increases significantly with velocity and in fact becomes unbound as the latter approaches $c$. One may thus construct a deterministic model of three fast rotating light particles which uses gravitation as the attractive force, special relativity, the weak equivalence principle and introduces angular momentum quantization as in the Bohr model of the H atom. The model shows the existence of stable rotational states corresponding to highly relativistic particle velocities with radii in the $fm$ range. When the rest masses of the three light particles is in the mass range of neutrinos $(\sim 0.1\;eV/c^2)$, then surprisingly the rest masses and Compton wavelengths of these rotational states are found to correspond to those of baryons and in fact the masses $(\sim 1\;GeV/c^2)$ reproduce quite well the light baryon mass spectrum. Interestingly the model also predicts a lower limit for the lifetime of the baryons which is consistent with their measured lifetimes and yields a good approximation for their magnetic moments. It thus appears that this deterministic Bohr-type approach to subatomic phenomena, i.e. a classical special relativistic treatment coupled with the de Broglie wavelength expression to seek conformity with quantum mechanics, may be a useful zeroth-order model to explore some quantum gravity problems.\\


\end{abstract}

\section{Introduction}
Quantum gravity seeks for the synthesis of general relativity and quantum mechanics \cite{Rovelli04}. The synthesis of special relativity and quantum mechanics still also remains to some extent an open problem \cite{Hooft07} It is anticipated that in general the synthesis of relativity and quantum mechanics can also shed new light on the structure and properties of the subatomic world, an area currently treated by quantum chromodynamics (QCD) \cite{Hooft07,Braun07}. 

On the other hand the Bohr and Bohr-Sommerfeld rotating electron models of the H atom present striking examples where classical  deterministic treatment of a physical problem coupled with the de Broglie wavelength expression leads to results in almost quantitative agreement with experiment and with the results of the full quantum mechanical Schr\"odinger treatment. In these very successful classical circular or elliptical orbit models the centripetal force is the Coulombic attraction between the proton and the electron which are both treated as point charges. 

It thus becomes interesting to examine the extent to which this deterministic type approach may be useful to provide a satisfactory zeroth order model for other physical problems at the subatomic level such as the condensation of the quark-gluon plasma to form hadrons \cite{Hooft07,Braun07}. In such an approach the parton or quark-gluon constituents of hadrons may be viewed as particles with relativistic velocities \cite{Nambu84,Schwarz04}.

Such an approach appears worthwhile to consider since, as analyzed in several recent papers by t' Hooft \cite{Hooft97,Hooft99,Hooft05,Hooft2005}, attempts to produce physically realistic models for the unification of general or special relativity with quantum mechanics may not be successful unless some level of determinism in the latter is restored \cite{Hooft99}. As shown by t' Hooft \cite{Hooft05} such models which introduce determinism beneath quantum mechanics and thus restore causality and locality can be constructed so that stochastic behaviour is correctly described by quantum mechanical amplitudes in precise accordance with the Copenhagen-Bohr-Born doctrine \cite{Hooft05}. A difficulty, however, with such models, which may be essential to restore causality and locality at Planckian distance scales \cite{Hooft05,Hooft2005}, is to obtain a Hamiltonian which is bounded from below and thus leads to a ground state \cite{Hooft05}. These considerations have provided an incentive for the present work where we present a deterministic model which combines special relativity, Newton's gravitational law and the weak equivalence principle to examine the rotational motion of three light neutral or charged particles under the influence of their gravitational attraction with velocities very close to the speed of light $c$. This leads to an infinity of bound rotational states and, similarly to the Bohr model of the H atom, angular momentum quantization is used to select among these states the ones which conform to the de Broglie wavelength expression which historically has been the origin of quantum mechanics.

Interestingly it is found that when the masses of the light particles are in the mass range of neutrinos $(\sim 0.1$ $eV/c^2)$ \cite{Griffiths08} then the masses of the confined composite bound states formed are in the hadrons mass range $(\sim 1\; GeV)$. Both the computed radii, which are in the $fm$ range, and the estimated magnetic moments are in good agreement with experiment. The model also provides a good approximation of the mass ratios between the proton and other baryons and also predicts a lower limit for baryon lifetimes $(\sim 10^{-23}\; s)$ which is in agreement with experiment. 

This deterministic Bohr-type hadronization model involves no unknown parameters and, although the observed very good agreement with experiment could be fortuitous, it appears to be a useful zeroth order model for hadronization.

It is at first quite surprising that this semiclassical deterministic model provides a good fit to experiment by considering gravity between the relativistic particles as the only attractive force. However, the possibility that gravity may have a significant role at short, femtometer or subfemtometer distances has attracted significant interest for years \cite{Hehl80,Hoyle01,Long03,Oldershaw10} and the potential role of special relativity \cite{Schwarz04} as well as the feasibility of developing a unified gauge theory of gravitational and strong forces \cite{Hehl80} have been discussed. Of special interest in this area was Wheeler's concept of geons (gravitational-electromagnetic entities), i.e. of spherical or toroidal objects made up of electromagnetic waves or neutrinos held together gravitationally \cite{Wheeler55,Misner73,Misner09}. Geons are classical entities based on general relativity. Small electromagnetic or neutrino geons had been proposed as classical models for elementary particles \cite{Wheeler55,Misner73,Misner09}.

More commonly the formation of baryons and other hadrons via condensation of smaller particles, i.e. of the gluon-quark plasma is analyzed in the QCD theory \cite{Braun07,Gross73,Politzer73,Cabibbo75}. This condensation occurs at the transition temperature of QCD which is estimated to be around 150 MeV in the kT scale, i.e. it corresponds to a particle energy of about 150 MeV \cite{Braun07}.

\section{Rotational model}
\subsection{Basic features and incentive}
The gravitational interaction is usually extremely weak and insignificant when discussing interaction between particles such as those forming baryons. In the model discussed here a baryon is described in terms of the rotational motion of three symmetrically spaced point particles with rest mass $m_o$ being significantly lower than $m_B$, the rest mass of the baryon. These three light point particles form a bound rotational state and are moving at a speed close to the speed of light, $c$. Both their kinetic energy and their inertial mass are significantly enhanced by this high relativitic speed. The rest mass $m_B$ of the baryon is determined by the total rest plus kinetic energy of the three particles, i.e. $m_B=3\gamma m_o$ where $\gamma(=(1-\texttt{v}^2/c^2)^{-1/2})$ is the Lorentz factor. When v approaches $c$ the Lorentz factor $\gamma$ and the inertial mass of the particle, $\gamma^3m_o$ \cite{French68,Freund08} increase very significantly. 

Thus the first incentive for the present work has been the observation that according to the equivalence principle between inertial and gravitational mass \cite{Schwarz04}, such a high value of the inertial mass implies an equally high value of the gravitational mass and thus, according to Newton's gravitational law, a much stronger gravitational attraction than that corresponding to non-relativistic velocities. Such a strong attraction can, as shown by the present model, lead to stable highly relativistic rotational states.

A second incentive has been the observation depicted in Fig. 1 and Table A1 in Appendix A that the masses of the known uncharmed spin (1/2) and spin (3/2) baryons \cite{Griffiths08}, i.e. of baryons consisting of u, d and s quarks, follow with good approximation, better than 3\%, a $(2n-1)^{1/6} $ law where $n=1$ corresponds to the proton or the neutron mass. This $(2n-1)^{1/6} $ law is reminiscent of the $n^{-2} $ law obeyed by the energy levels of the H atom. 

In the model discussed here the three light particles are assumed to have a spin 1/2. They are also allowed to be charged which allows for charged baryons, such as the proton. However, the Coulomb interaction between them is shown to contribute less than 0.1 percent to the composite (baryon) mass. The Coulomb interaction seems only to account for small differences between close masses of baryons e.g. between that of the proton and the neutron.
\begin{figure}[ht]
\begin{center}
\includegraphics[bb=0mm 0mm 208mm 296mm, width=75.6mm, height=65.8mm, viewport=3mm 4mm 205mm 292mm]{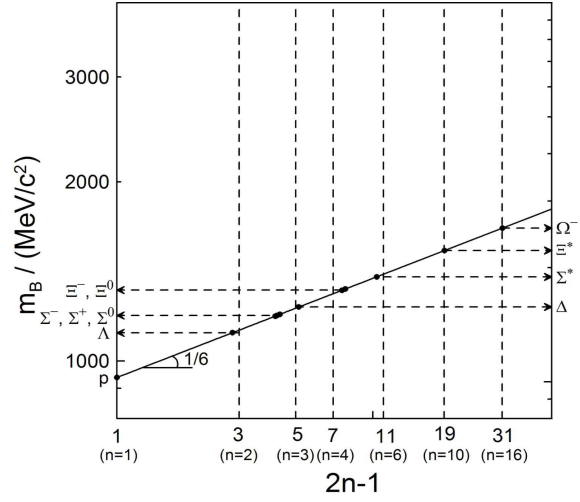}
\end{center}
\caption{Comparison of the masses, $m_B$, of the uncharmed baryons, consisting of u, d and s quarks, \cite{Griffiths08} with equation \eqref{eq23}, i.e. $m_B=m_p(2n-1)^{1/6}$ where $m_p$ is the proton mass.}
\label{fig:1}
\end{figure}

\subsection{Equivalence principle and the gravitational force under relativistic conditions}
We consider a laboratory frame $S$ and an instantaneous inertial frame $S'$ moving with a particle with an instantaneous velocity  $\textbf{v}$ relative to frame $S$. 

It is worth noting that the instantaneous inertial frame $S'$ is defined by the vector $\textbf{v}$ alone and not by the overall type of motion (e.g. linear or cyclic) performed by the particle \cite{French68}.

Thus one can assign to the frame $S'$ and corresponding velocity $\textbf{v}$ an inertial particle mass, $m_i$, by considering a test force, $\textbf{F}$, parallel to $\textbf{v}$, acting on the particle. According to the theory of special relativity the case where $\textbf{F}$ and $\textbf{v}$ are parallel is the only case where the force is invariant, i.e. the force perceived in $S$ and $S'$ is the same \cite{French68}. 

Starting from the general relativistic equation of motion, i.e. from \cite{French68,Freund08}:
\begin{equation}
\label{eq1n}
\textbf{F}=\frac{d\textbf{p}}{dt}=\gamma m_o\frac{d\textbf{v}}{dt}+\gamma^3m_o\frac{1}{c^2}\left(\textbf{v}\cdot\frac{d\textbf{v}}{dt}\right)\textbf{v}
\end{equation}
and using the fact that the test force $\textbf{F}$ is taken to be parallel to $\textbf{v}$ one obtains after some simple algebra that the measures of the force, $F$, and of the acceleration, {dv/dt}, are related via:
\begin{equation}
\label{eq2n}
F=\frac{dp}{dt}=\frac{d(\gamma m_o\texttt{v})}{dt}=\gamma^3m_o\frac{d\texttt{v}}{dt}
\end{equation}

This defines the mass $\gamma^3m_o$, termed sometimes longitudinal mass \cite{Freund08}, which is the inertial mass of the particle, $m_i$, as it equals the ratio of force and acceleration. Thus it is $m_i=\gamma^3m_o$. According to the equivalence principle, $m_i$ also equals the gravitational mass, $m_g$, of the particle \cite{Schwarz04}, i.e.:
\begin{equation}
\label{eq3n}
m_g=m_i=\gamma^3m_o
\end{equation}

It is worth noting that the equivalence principle has been formulated and tested for parallel $\textbf{F}$ and $\textbf{v}$ vectors, as is the present case. It is also worth reminding that, as already noted, for given $m_o$ and v the inertial mass $m_i$ and thus the gravitational mass $m_g$ are both uniquely determined by \eqref{eq2n} and their value does not depend on the type of motion (e.g. linear or circular) performed by the particle. 

Thus upon considering a second particle of rest mass $m_o$ and instantaneous velocity measure v relative to the observer at $S$ and at a distance $r$ from the first particle, it follows that the inertial and gravitational mass of the second particle is also given by $\gamma^3m_o$, as in equation \eqref{eq3n} and thus one can use these two $m_g$ values in Newton's gravitational law in order to compute the gravitational force, $F_G$, between the two particles. One thus obtains:
\begin{equation} 
\label{eq2} 
F_{G}=-\frac{Gm_{o}^{2} \gamma ^{6}}{r^{2}}
\end{equation}
which depends on the $6^{th}$ power of $\gamma$ \cite{Vayenas09}. It is worth remembering that this equation stems directly from special relativity (eq. \ref{eq2n}), the weak equivalence principle (eq. \ref{eq3n}) and Newton's gravitational law.

In our semi-classical model the three interacting particles composing the baryon, are assumed to rotate at the same speed around their center of mass, in a circular orbit of radius $R$, as shown in Fig. 2. For the laboratory observer the tangential velocity v is close to the speed of light, $c$ and the center of mass of the three particles is at rest. The magnitude of the resulting force acting on each particle is:
\begin{equation}
\label{eq3}
F_{G}=-\frac{Gm_{o}^{2} \gamma ^{6}(R)}{\sqrt{3} {\rm \;}R^{2}}
\end{equation}
\begin{figure}[ht]
\begin{center}
\includegraphics[bb=0mm 0mm 208mm 296mm, width=69.6mm, height=88.0mm, viewport=3mm 4mm 205mm 292mm]{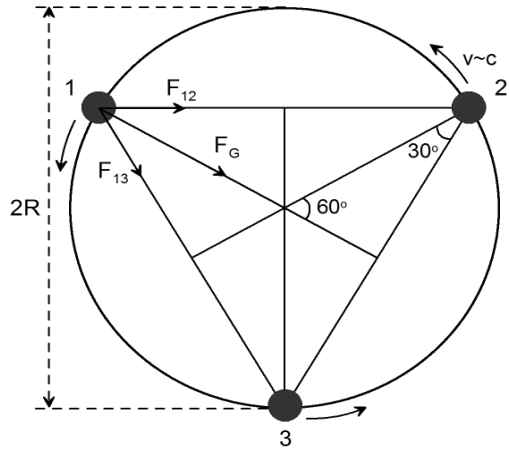}
\end{center}
\caption{Three particles moving at a constant tangential velocity, $v$, in a circle of radius $R$ around their center of mass. They are equally spaced. $F_{12}$ and $F_{13}$ are two particle attraction forces and $F_{G}$ is the resultant, radial, force.}
\label{fig:2}
\end{figure}
\section{Rotational bound states}
\subsection{Centripetal force}
One may now consider the relativistic equation of motion (eq. \eqref{eq1n}) for the case of a circular orbit of radius $R$ with $F=F_G$. Since $\textbf{v}\cdot\frac{d\textbf{v}}{dt}=0$ one obtains:
\begin{equation} 
\label{eq4n} 
\textbf{F}_G=\gamma m_o\frac{d\textbf{v}}{dt}=\gamma m_o\frac{\texttt{v}^2}{R} \qquad and \qquad F_G=\gamma m_o \frac{\texttt{v}^2}{R}
\end{equation}
and thus upon combining with eq. \eqref{eq3} one obtains:
\begin{equation} \label{eq4} 
\frac{Gm_{o}^{2}\gamma^{6}(R)}{\sqrt{3}{\rm \; }R^{2}}=\frac{\gamma (R)m_{o}\texttt{v}^{2}}{R}
\end{equation}
i.e., the gravitational force $F_G$ given by \eqref{eq3} acts as the centripetal force for the rotational motion. 

It must be noted on the basis of \eqref{eq4n} one might be tempted to assign the value $\gamma m_o$, commonly termed transverse mass \cite{French68,Freund08}, to the inertial and thus gravitational mass of each particle. However, as already noted the mass $m_i(=m_o\gamma^3)$, and thus also $m_g$ is uniquely determined for given $m_o$ and $\textbf{v}$, via the colinear to $\textbf{v}$ test force $\textbf{F}$, and does not depend on the type of motion (e.g. linear or circular) performed by the particle.

Upon utilizing $\gamma (R)=(1-\texttt{v}^{2} /c^{2} )^{-1/2}$ in \eqref{eq4} one obtains:
\begin{equation} \label{eq5} 
R=\frac{Gm_{o}}{\sqrt{3}{\rm \; c}^{{\rm 2}} } \gamma ^{5} \left(\frac{\gamma ^{2}}{\gamma^{2}-1}\right)
\end{equation}
or, equivalently:
\begin{equation} \label{eq6} 
R=(R_{S}/(2\sqrt{3}))\gamma ^{5} \left(\frac{\gamma ^{2}}{\gamma ^{2}-1} \right)
\end{equation}
where $R_{S}(=2Gm_{o}/c^{2})$ is the Schwarzschild radius of a particle with rest mass $m_o$.
\begin{figure}[ht]
\begin{center}
	\includegraphics[bb=0mm 0mm 208mm 296mm, width=72.9mm, height=68.6mm, viewport=3mm 4mm 205mm 292mm]{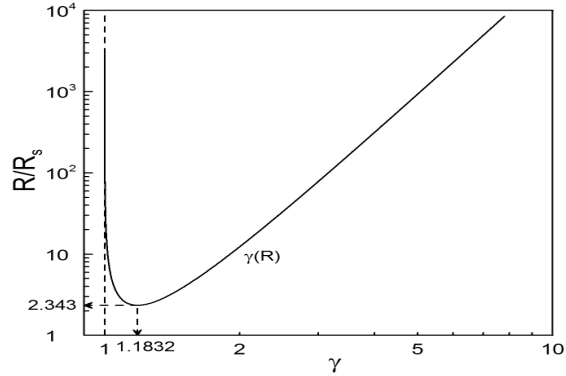}
	\end{center}
	\caption{Plot of eq. \eqref{eq6} near the minimum R.}
	\label{fig:3}
\end{figure}

As shown in Figure 3 the function R defined by eq. \eqref{eq6} exhibits a minimum, $R_{\min}=2.343R_{S} $, at $\gamma _{\min}=\sqrt{7/5} =1.1832$ thus v$_{\min}=\sqrt{2/7}c$. This is the minimum radius for a circular orbit and the corresponding minimum angular momentum is $L_{\min}=\gamma _{\min}m_{o}v_{\min}R_{\min}=1.481m_{o}cR_{S}=2.963$ $Gm_{o}^{2}/c$. This condition is similar to the criteria $L>Gm^2/c$ or $R>R_{S} /2$ found for circular orbits in special relativity \cite{Torkelsson98,Boyer04} or $L>2\sqrt{3} Gm^{2} /c$ for the Schwarzschild metric in general relativity \cite{Boyer04} with orbits around point masses with $R^{-1}$ potentials.

Equation \eqref{eq6} defines two $\gamma $ branches (Fig. 3), one corresponding to low $\gamma$ values $(\gamma <1.1832)$ the other corresponding to large $\gamma$ values $(\gamma > 1.1832)$. The first branch corresponds to common gravitational orbits. In this case $\gamma$ and thus the velocity $v$ decreases with increasing R, e.g. $\texttt{v}=(Gm_{o}/(\sqrt{3}R))^{1/2}$ in the non-relativistic case $(\gamma \approx 1)$.

The second branch which leads to relativistic velocities defines rotational states where $\gamma$ and thus v increase with increasing R. These states with $\gamma \gg 1$ are the states of interest to the present model. For $\gamma \gg 1$, e.g. $\gamma >10^{2} $ eq. \eqref{eq6} reduces to:
\begin{equation} \label{eq7}
R=(R_{S}/(2\sqrt{3}))\gamma^{5}\quad;\quad \gamma=(2\sqrt{3}R/R_S)^{1/5}
\end{equation}

This equation can be used to eliminate R or $\gamma $ in the force expression of eq. \eqref{eq3}. In the former case (i.e. elimination of R) one obtains:
\begin{equation} \label{eq8}
F_{G}=-\frac{\sqrt{3} c^{4}}{\gamma^{4}G}
\end{equation}
and thus, interestingly, for any fixed value of $\gamma$ or $R$, the attractive force is uniquely determined by the familiar $G/c^4$ parameter of the gravitational field equations of general relativity \cite{Wheeler55,Misner73}, i.e. $G_{ik}=(8\pi G/c^4)T_{ik}$, which relates the Einstein tensor $G_{ik}$ with the stress-momentum-energy tensor $T_{ik}$ \cite{Wheeler55,Misner73}.

In the latter case, i.e. elimination of $\gamma $, one obtains:
\begin{equation} \label{eq9} 
F_{G} =-m_{o} c^{2} \left(\frac{2\sqrt{3} }{R_{S} } \right)^{1/5} \frac{1}{R^{4/5} }
\end{equation}

\subsection{Potential, translational and total energy}
The force equation \eqref{eq9} refers to circular orbits only and thus defines a certain conservative force, since the work done in moving the particles between two points $R_1$ and $R_2$, corresponding to two rotational states with radii $R_1$ and $R_2$, is independent of the path taken. The force vector is always pointing to the center of rotation and thus a conservative vector field is defined which is the gradient of a scalar potential, denoted $V_G(R)$. The latter can be termed gravitational potential energy of the three rotating particles and corresponds to the transfer of the particles from the minimum circular orbit radius $R_{\min }$ to an orbit of radius of interest, $R$. The function $V_G(R)$ is obtained via integration of eq. \eqref{eq9}, i.e., denoting by $R'$ the dummy variable:
\begin{eqnarray}\label{eq11}
V_{G}(R)-V_{G}(R_{min})&=&\int_{R_{min}}^{R}F_{G}dR'=\\&=&
\nonumber -5m_{o}c^{2} \left(\frac{2\sqrt{3}}{R_{S}}\right)^{1/5}\left(R^{1/5}-R_{\min}^{1/5}\right)
\end{eqnarray}

Thus while the magnitude of the gravitational force acting on the rotating particles increases with decreasing radius, R (eq. \eqref{eq9}), the absolute value $\left|V_G(R)\right|$ of the gravitational potential energy increases with increasing $R$. This behavior is reminiscent of asymptotic freedom \cite{Hooft07,Gross73,Politzer73,Cabibbo75}, i.e. the attractive interaction energy is small at short distances and increases significantly with increasing distance $R$. 

Since it is $\gamma _{\min } \approx 1$, the value of $V_{G}(R_{\min})$ can be estimated from Newton's gravitational law and is about $-3.338$ $m_{o} c^{2}$, which, as shown in section 3.5, turns out to be of the order of 0.1 eV. This is quite negligible in comparison with the rest energies of baryons $(\sim GeV)$ of interest in this work. Thus, in this scale, $V_{G} (R_{\min})$ is quite close to the zero gravitational potential energy of the three free particles with zero velocity at infinite distance from each other. Therefore when this vacuum state is chosen as the reference gravitational potential energy state $(V_{G}=0)$ then for $R\gg R_{\min } $ eq. \eqref{eq11} is written with very good accuracy as:
\begin{equation} \label{eq12}
V_{G}(R)=-5m_{o}c^{2}\left(\frac{2\sqrt{3}}{R_{S}}\right)^{1/5}R^{1/5}
\end{equation}
or using Eq. \eqref{eq7}:
\begin{equation} \label{eq13}
V_{G}(R)=-5\gamma m_{o} c^{2}
\end{equation}

On the other hand the kinetic energy, $T$, of the three rotating particles is:
\begin{equation} \label{eq10}
T(R)=3(\gamma-1)m_{o}c^{2}
\end{equation}

Thus one may now compute the change in total system energy, $\Delta E$, upon formation of the rotational bound state from the three originally free particles. Denoting by f and i the final and initial states and by (RE) the rest energy, one obtains from energy conservation:
\begin{eqnarray} \label{eq14}
\Delta E&=&E_{F}-E_{i}=\\
\nonumber &=&\left[(RE)_{f}+T_{f}+V_{G,f}\right]-\left[(RE)_{i}+T_{i}+V_{G,i} \right]=\\
\nonumber &=&\left[3m_{o}c^{2}+3(\gamma -1)m_{o}c^{2}-5\gamma m_{o}c^{2}\right]-3m_{o} c^{2}=\\ 
\nonumber &=&\Delta T+\Delta V_{G} =-(2\gamma +3)m_{o} c^{2} \approx -2\gamma m_{o} c^{2}
\end{eqnarray}
where the last equality holds for $\gamma \gg 1$ as is the case of interest here.

The same $\Delta E$ expression is, of course, obtained regardless of the choice of the reference potential energy state. Thus in view of eqs. \eqref{eq7}, \eqref{eq12}, \eqref{eq13} and \eqref{eq14} one can summarize the dependence of $\Delta T$, $\Delta V_G$ and $\Delta E$ on $\gamma$ and $R$ for $\gamma>>1$ as: 
\begin{equation} \label{eq15} 
\Delta T=T=3(\gamma -1)m_oc^2\approx 3\gamma m_oc^2=3m_oc^2\left(\frac{2\sqrt{3}R}{R_S}\right)^{1/5}
\end{equation}
\begin{equation} \label{eq16} 
\Delta V_G=-5\gamma m_oc^2=-5m_oc^2\left(\frac{2\sqrt{3}R}{R_S}\right)^{1/5}
\end{equation}
\begin{equation} \label{eq17} 
\Delta E=-(2\gamma +3)m_oc^2\approx -2\gamma m_oc^2=-2m_oc^2\left(\frac{2\sqrt{3}R}{R_S}\right)^{1/5}
\end{equation}

The negative sign of $\Delta E$ denotes that the formation of the bound rotational state starting from the three initially free particles happens spontaneously, is exoergic $(\Delta E<0)$, and the binding energy $BE(=-\Delta E)$ equals $2\gamma m_oc^2$.

\subsection{Composite particle rest energy and binding energy}
The total rest plus kinetic energy of the three rotating particles equals $3\gamma m_oc^2$ and constitutes at the same time the rest energy of the composite particle formed, i.e. of the rotational bound state. Denoting by $m$ the rest mass of the composite bound state, it is:
\begin{equation} \label{eq18} 
mc^2=3\gamma m_oc^2
\end{equation}

It is useful to note that in the model the rest mass of the three particles, i.e. $3m_oc^2$, does not change when the bound state is formed. The transformation of the kinetic energy of the three rotating particles into rest energy of the bound state, is simply due to the change in choice of the boundaries of the system. In the former case (three individual rotating particles) the boundaries are geometrically disconnected and encompass each particle individually, in the latter case the system boundary contains all three particles. Thus the formation of the bound rotating state by the three particles provides a simple hadronization mechanism, i.e. generation of rest mass, $m$, starting from an initial rest mass $3m_o$, according to eq. (\ref{eq18}).

It follows from \eqref{eq14} and \eqref{eq18} that:
\begin{equation} \label{eq19}
BE=-\Delta E=(2/3)mc^{2}
\end{equation}

Thus the binding energy per light particle is $(2/9)mc^2$, which for $m=m_p=938.272\;MeV/c^2$, the proton mass, gives an energy of 208 $MeV$, not far from the estimated particle energy of 150 $MeV$ at the transition temperature of QCD \cite{Braun07}.

One may note here that since the potential energy expression \eqref{eq13} does not depend on the number, $N$, of rotating particles but the kinetic energy, T, does, i.e. $N(\gamma -1)m_oc^2$, it follows from \eqref{eq14} that, according to the model, stable rotational states cannot be obtained for $N>5$ since they lead to positive $\Delta E$. The case $N=2$ is interesting, as it may be used to model mesons, and gives results in qualitative agreement with those obtained for the $N=3$ case treated here. For $N=2$ one finds that the computed composite masses, $m$, are in the $0.5\;GeV/c^2$ range.

\subsection{Quantization of angular momentum}
We then proceed to identify among the infinity of bound rotational states described by eqs. \eqref{eq15}, \eqref{eq16} and \eqref{eq17}, each corresponding to a different $R$, those rotational states where $R$ is an integer multiple of the reduced de Broglie wavelength\; ${\mathchar'26\mkern-10mu\lambda}(=\hbar /p)$ of the light rotating particles.

This can be done by introducing quantization of the angular momentum of the light particles in the form:
\begin{eqnarray} \label{eq20}
L=\gamma_{B}m_{o}R_B c=(2n-1)\hbar {\rm \; \; };\\
\nonumber R_{B}=(2n-1)\frac{\hbar }{\gamma_{B} m_{o}c}=(2n-1)\frac{3\hbar}{m_{B}c}
\end{eqnarray}
where the subscript $B$ denotes a baryon and indicates that the corresponding $\gamma$, $R$ and $m(=3\gamma m_o)$ values satisfy the quantization condition \eqref{eq20}.

It is assumed that $n=1$ corresponds to a proton and thus eq. \eqref{eq20} yields, for $n=1$, $R_B(1)=3\hbar/m_pc$ which equals the reduced de Broglie wavelength of the light particle of rest mass $m_o$ moving at a speed $\texttt{v}\approx c$ and is three times the reduced Compton length of the proton, ${\mathchar'26\mkern-10mu\lambda}_{c,p}(=\hbar /m_p c)$.

The mass $m_{B}$ is calculated by combining Eqs. \eqref{eq7}, \eqref{eq18} and \eqref{eq20}:
\begin{equation} \label{eq21}
m_{B}=3^{13/12}(2n-1)^{1/6} m_{o}^{2/3} \left(\frac{\hbar c}{G} \right)^{1/6}
\end{equation}

One notes that $(\hbar c/G)^{1/2}$ is the Planck mass, $m_{Pl}$ \cite{Wheeler55}. One can then reasonably assume that the ground state $(n=1)$ corresponds to a proton (and neutron). This yields quantitative expressions for the proton mass and for $m_{o} $:
\begin{equation} \label{eq22}
m_{p}=3^{13/12}m_{o}^{2/3}\left(\frac{\hbar c}{G} \right)^{1/6}
\end{equation}
where $m_p$ is the proton mass $(938.272\; MeV/c^2)$. The mass of the heavier baryons follow the series:
\begin{equation} \label{eq23}
m_{B}=m_{p}(2n-1)^{1/6}
\end{equation}
which is in good agreement with the experimental data as shown in Figure 1 and Table A1.

\subsection{Computation of $m_o$, $\gamma_B$ and $R_B$}
From eq. \eqref{eq22} one obtains the following expression for $m_{o}$:
\begin{equation} \label{eq24}
m_{o}=\frac{(m_{p}/3)^{3/2}}{3^{1/8}\left(\frac{\hbar c}{G}\right)^{1/4}}
\end{equation}
Substituting the values of the constants one obtains:
\begin{equation} \label{eq25} 
m_{o}=7.777\times 10^{-38} kg=0.0436{\rm \; eV/c^2}
\end{equation}
which, surprisingly, is in the mass range of neutrinos \cite{Griffiths08}. The heaviest neutrinos mass is currently estimated to be between 0.04 and 0.4 eV/c$^2$ \cite{Griffiths08}.

The value of $\gamma _{B} $ for the proton, $\gamma_{B}(1)$, is obtained from Eqs. \eqref{eq18}, \eqref{eq25} i.e.:
\begin{equation} \label{eq26}
\gamma_{B}(1)=3^{1/12}\left(\frac{\hbar c}{G} \right)^{1/6}/m^{1/3}_o=7.169\times 10^{9}
\end{equation}
while for the heavier baryons it increases with n,
\begin{equation} \label{eq27}
{\rm \gamma }_{{\rm B}} {\rm (n)}=7.169\times 10^{9} (2n-1)^{1/6}
\end{equation}

The rotational radius R for the baryons increases with n, as given by Eqs. \eqref{eq20} and \eqref{eq27}:
\begin{equation} \label{eq28}
R_{B{\rm \; }} (n)=(2n-1)^{5/6} R_{B} (1)=(2n-1)^{5/6} 6.31\cdot 10^{-16} \; m
\end{equation}
where $R_{B}(1)=R_{p}=6.31\cdot 10^{-16} \; m$ corresponds to the proton.
\begin{figure}[ht]
\begin{center}
\includegraphics[bb=0mm 0mm 208mm 296mm, width=70.6mm, height=95.1mm, viewport=3mm 4mm 205mm 292mm]{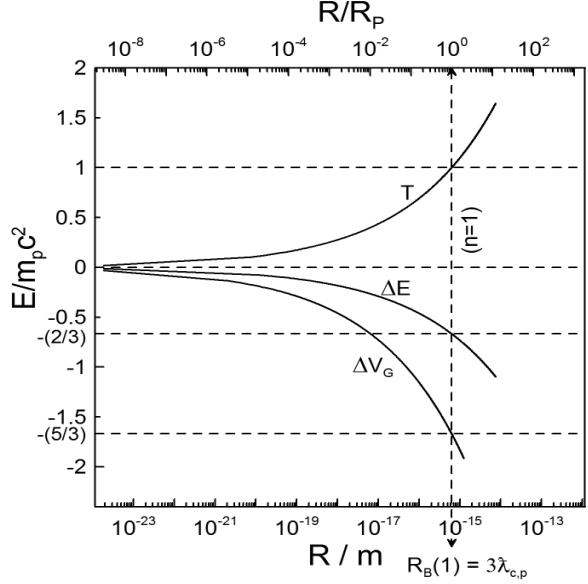}
\end{center}
\caption{Plot of eq. \eqref{eq20} for $n=1$ (vertical line) and of eqs. \eqref{eq29}, \eqref{eq30}, \eqref{eq31}, showing how the Bohr quantization condition (eq. \eqref{eq20}) determines the rest energy, $m_pc^2(\approx T)$ of the bound state and the corresponding binding energy $-\Delta E$.}
\label{fig:4}
\end{figure}
Using the definitions of $R_B(1)=R_p(=3\hbar/m_pc)$, of $R_S(=2Gm_o/c^2)$ and eq. \eqref{eq18}, i.e. $m_p=3\gamma m_o$ one can rewrite eqs. \eqref{eq15}, \eqref{eq16} and \eqref{eq17} in terms of $R_p$ rather than $R_s$:
\begin{equation}
\label{eq29}
\Delta T=T=m_pc^2(R/R_p)^{1/5}
\end{equation}
\begin{equation}
\label{eq30}
\Delta V_G=-(5/3)m_pc^2(R/R_p)^{1/5}
\end{equation}
\begin{equation} \label{eq31}
\Delta E=-(2/3)m_pc^2(R/R_p)^{1/5}
\end{equation}
Plots of these functions are given in Fig. 4.

\subsection{Rotational periods}
The period of rotation $\tau(n)$ of the light particles within the composite ones, $2\pi R/\texttt{v}\sim 2\pi R/c$, is, using Eq. \eqref{eq28},
\begin{equation} \label{eq32}
\tau (n)=(2n-1)^{5/6} \tau _{p}=(2n-1)^{5/6} 6.6\cdot 10^{-24} \;s
\end{equation}
where $\tau_{p}=2\pi R_{p}/c=6.6\times 10^{-24} \;s$ is the rotation period for the proton.  The time interval $\tau(n)$ provides a rough lower limit for the lifetime of the composite particles, interpreted as baryons, as they can be defined only if the light particles complete at least a revolution (Fig. 2). Indeed all the known lifetimes of the baryons are not much shorter than that estimate. The lifetime of the $\Delta $ baryons, which is the shortest, is $5.6\cdot 10^{-24} s$ \cite{Griffiths08}.

\section{The Coulomb interaction}
So far the model discussed only the mass of the light particles and its effect on the binding forming a neutral composite particle. However, baryons  also possess a spin and many of them also charge. We therefore allocate to the light particles, in analogy to the quark model \cite{Hooft07,Nambu84}, a charge and spin. It is assumed that there are two types of light particles, those with charge $2e/3$ and those with $-e/3$, where e is the elementary charge. In addition each light particle is assumed to possess a spin 1/2. As a result all composite particles have either a total spin 1/2 or 3/2 as is the case for most baryons.

The Coulombic forces between charged particles with relativistic velocities have been studied in detail \cite{French68}. It is well established that Coulomb's law correctly gives the force on the test charge for any velocity of the test charge provided the source charge is at rest \cite{French68}. In the simplified geometry of Figure 2 the distance between the two particles remains constant, thus in the reference frame of the source charge the test charge is also at rest, thus Coulomb's law remains valid without any relativistic corrections. 

It is thus possible to estimate the Coulomb interaction energy between the rotating particles. In the simplified geometry of Figure 2 the total Coulomb potential energy for the proton (charges 2/3, 2/3, -1/3) vanishes (since (4/9)-(2/9)-(2/9)=0), while for the neutron (charges -1/3, -1/3, 2/3) it is negative, i.e. denoting $\varepsilon =4\pi \varepsilon_{o}$ one obtains:
\begin{equation} \label{eq33}
V_{C,n}=\frac{e^{2}}{\varepsilon \sqrt{3} R_{B} } \left[(1/9)-(2/9)-(2/9)\right]=-\frac{(e^{2} /\varepsilon )}{3\sqrt{3} R_{B}}
\end{equation}
, i.e. there is an overall attractive Coulombic interaction.

Upon substituting $R_{B} $ from Eq. \eqref{eq20} for the case of the proton $(n=1)$ one obtains:
\begin{equation} \label{eq34}
V_{C,n} =-\frac{e^{2} }{9\sqrt{3} \varepsilon c\hbar } m_{p} c^{2}=-\frac{\alpha }{9\sqrt{3} } m_{p} c^{2}
\end{equation}
where $\alpha (=e^{2} /\varepsilon c\hbar=1/137.0359) $is the fine structure constant. Thus $V_{C,n} $ is about three orders of magnitude smaller than the rest energy of the proton. This justifies the approximation made by neglecting the Coulomb interaction.

Interestingly as shown in Table A2 (Appendix A) the mass differences in baryons which differ only in the charge value (e.g. the N, $\Sigma $ or $\Xi $ baryons) are generally small (up to 7 $MeV/c^2$) and the ratio $\Delta m_{N} /m_{N} $, $\Delta m_{\Sigma } /m_{\Sigma } $ or $\Delta m_{\Xi } /m_{\Xi } $ is of the order of $10^{-3}$, similarly to the value of the ratio $(V_{C,n} /c^{2} )/m_{p} $ obtained from eq. \eqref{eq34}. Thus the Coulomb interaction could be the origin of this small difference.

If the Coulomb interaction is taken into consideration the symmetry of the configuration of Fig. 2 is broken as not all three charges are the same. Although the deviation from three-fold symmetry is small, since the Coulombic energy is small, and thus one may still use with good accuracy eq. \eqref{eq33} to estimate the attractive interaction between the three particles forming a neutron, it is conceivable that this broken symmetry may be related to the relative instability of the neutron (lifetime 885.7 $s$) vs the proton (estimated lifetime $\sim 10^{30} s$ \cite{Griffiths08}).

\section{Magnetic moments}
It is interesting to compute the magnetic dipole moments, $\mu $, of these bound rotational states. Using the definition of $\mu (=(1/2)qRv)$ and considering the case $n=1$, corresponding to a proton (which is a uud baryon) with charge $2e/3$ for u and $-e/3$ for d it is:
\begin{equation} \label{eq35}
\mu_{p}=(1/2)eRc\left[(2/3)+(2/3)-(1/3)\right]=(1/2)eRc
\end{equation}

Upon substituting $R=R_{p}=0.631 \; fm$ one obtains:
\begin{equation}
\label{eq36}
\mu_{p}=15.14\cdot 10^{-27} {\rm \; J/T\; \; \; \; \; (=3\mu_N)}
\end{equation}
where $\mu _{N} $ is the nuclear magneton $(5.05\cdot 10^{-27} {\rm \; }J/T)$. This value differs less than 8\% from the experimental value of $14.10\cdot 10^{-27} {\rm \; J/T\; \; (i.e.\; 2.79\; }\mu _{{\rm N}} )$ \cite{Gillies97,Povh06}.

In the above computation (eq. \eqref{eq35}) one assumes that the spin vectors of the three small particles (i.e. uud) are parallel to the vector of rotation of the rotating proton state. If one considers the neutron which is a udd particle and assumes that the spin of one of the two d quarks is parallel with the rotation vector of the rotating neutron state and the spins of the other two particles are antiparallel to the neutron rotation vector then one obtains:
\begin{equation}
\label{eq37}
\mu_{n}=(1/2)eRc\left[(-2/3)+(1/3)-(1/3)\right]=-(1/3)eRc
\end{equation}
and upon substitution from Eq. \eqref{eq28} of $R=0.631\; fm$ one obtains:
\begin{equation} \label{eq38}
\mu_{n}=-10.09\cdot 10^{-27}{\rm \; J/T}=-2\mu_{N}
\end{equation}
which is in excellent agreement with the experimental value of $-9.66\cdot 10^{-27} {\rm \; J/T\; }(=-1.913\mu _{N} )$.

This good agreement seems to imply that the spin contribution of the light particles to the magnetic moment of the rotating state is small and only the spin vector orientation (parallel or antiparallel to the baryon rotation vector) is important.

\section{Discussion}
A simple semi-classical model is presented which explores the use of gravitation as the origin of the strong force by examining the rotational motion of three light particles due to their gravitational attaction under relativistic conditions. The model uses special relativity and the equivalence principle to express the gravitational force by substituting the inertial mass $\gamma ^{3}m_{o}$ in Newton's gravitational law. Interestingly for any fixed $\gamma$ value the gravitational force is found to be uniquely determined by the key parameter, $G/c^4$, of the field equations of general relativity. The model is similar to the original Bohr model of the H atom. In the present case the angular momentum is assumed to be quantized with the condition $(2n-1)\hbar$ rather than with $n\hbar$ as in the Bohr model for the H atom. 

The model provides surprisingly interesting results concerning the masses that follow a $(2n-1)^{1/6} $ law, the binding energies per light particle, and also the radii and the magnetic moments. It also provides a reasonable lower limit for the baryon lifetimes.

The mass $m_{o}=0.0436{\rm \; eV/c^2}$ and spin (1/2) of the light particles, agrees with that of the neutrino \cite{Griffiths08}. Interestingly this is consistent with Wheeler's concept of neutrino geons \cite{Wheeler55}. If this is the case then some of these hadron forming neutrinos have to be charged (at least in charged baryons). One may explain their absence in experiment by a high cross section for absorption, e.g. to form neutral neutrinos or hadrons.

As in the standard model, the charge allocated to the light particles is $(2/3)e$ and $-e/3$. The Coulomb interaction is typically a factor of $\alpha$ smaller than the gravitational one and thus can be as a first step neglected. It is of the order of the small difference in mass between the proton and the neutron. In the series $m_{B}=m_{p}(2n-1)^{1/6}$ baryons with n=5, 7, 8 and 9 are missing. This raises the possibility that such baryons may exist.

The $\gamma_{B}$ value for the proton is found to be $\gamma_B(1)=7.169\times 10^{9}$, which when inserted into Eq. \eqref{eq2} enhances the gravitational force by a factor of $\gamma_{B}^{6}\sim 10^{58}$. This is consistent with the initial hypothesis that the gravitational energy can be enhanced to the level of $m_Bc^2$. Interestingly it follows from equation \eqref{eq26} that in the case of the proton $(n=1)$ the inertial and gravitational mass of each rotating particle, $\gamma^3_Bm_o$, is related to the Planck mass, $m_{Pl}=(\hbar c/G)^{1/2}$, via a very simple equation, i.e. 
\begin{equation}
\label{eq39}
\gamma^3_B m_o=3^{1/4}m_{Pl}=3^{1/4}\left(\frac{\hbar c}{G}\right)^{1/2}=1.607\cdot 10^{19}\quad GeV/c^2
\end{equation}
which provides an interesting direct connection between the Planck mass and the present model. Interestingly, in Wheeler's geon analysis the minimum mass of a small geon also lies in the Planck mass range \cite{Wheeler55}. Gravity is generally expected to reach the level of the strong force at energies approaching the Planck scale $(\sim 10^{19}\;GeV)$ \cite{Schwarz04} which is in good agreement with the model results (eq. \eqref{eq39}).

It is interesting to note that when using the inertial or gravitational mass, $\gamma^3_Bm_o$, in the definition of the Compton wavelength, $\lambda_c$ of the particle $(=h/mc)$ then one obtains the Planck length $(\sim 10^{-35}\;m)$, but when using the mass corresponding to the total energy of the particles, $3\gamma_B m_o$, then one obtains the proton Compton wavelength $(\sim 10^{-15}\;m)$, which is close to the actual distance between the rotating particles.

Another interesting feature of the model is that it is qualitatively consistent with a central experimental observation about the strong force \cite{Hooft07}, i.e. that the normalized angular momentum of practically all hadrons and their excited states is roughly bounded by the square of their mass measured in $GeV$ \cite{Hooft07}. Indeed from \eqref{eq20} and \eqref{eq23} one obtains:
\begin{equation}
\label{eq40}
(L/\hbar)/(m_B/GeV)^2=1.13(2n-1)^{2/3}
\end{equation}
which in reasonable qualitative agreement with experiment for small integer $n$ values.

It thus appears that deterministic Bohr-type models which combine special relativity with the de Broglie wavelength equation may provide a useful zeroth order approach to explore some subatomic problems. The present model suggests that gravity may perhaps provide a mechanism for the binding of light particles, probably neutrinos, to form baryons. This is qualitatively consistent with Wheeler's concept of small neutrino geons \cite{Wheeler55}.

While the results are interesting they should be examined in light of the assumptions made, i.e. that (a) the force of gravitation for relativistic particles can be expressed by Eq. \eqref{eq2}, obtained by replacing via the equivalence principle the gravitational mass by the inertial mass $\gamma^3m_o$ in Newton's gravitational law, and (b) that $(2n-1)\hbar$ rather than $n \hbar$ can be used for the quantization of the angular momentum of these rotational states.
\section*{Acknowledgements}
CGV acknowledges numerous helpful discussions with Professor Ilan Riess of the Physics Department at the Technion.

\newpage
\section*{APPENDIX A}
\begin{table}[ht]
\caption{Experimental \cite{Griffiths08} and computed (eq. \eqref{eq23}) baryon masses}
\begin{tabular}{|p{0.35in}|p{0.7in}|p{1.5in}|p{1.00in}|p{0.3in}|p{0.5in}|}\hline
Baryon & Quark Content & Experimental mass value $MeV/c^2$ & $m_p(2n-1)^{1/6}$ & n & 2n-1\\ \hline
\vspace{0.1cm}$N \left\{\begin{array}{l}{p}\\{n}\end{array}\right.$ & \vspace{0.1cm}uud\newline udd &\vspace{0.1cm} 938.272\newline 939.565 & \vspace{0.1cm}938.272 &\vspace{0.1cm} 1 &\vspace{0.1cm} 1\\ \hline 
$\Lambda$ & uds & 1115.68 & 1126.8 & 2 & 3 \\ \hline
\vspace{0.1cm}$\Sigma^{+}$\newline $\Sigma^{o}$\newline $\Sigma^{-} $\newline $\Delta$ &\vspace{0.1cm} uus\newline uds\newline dds\newline uuu,uud, udd,ddd &\vspace{0.1cm} $\left.\begin{array}{l}1189.37\\1192.64\\1197.45\\{}\\1232\end{array}\right\}$\vspace{0.1cm} &\vspace{0.8cm}1226.9 & \vspace{0.6cm}3&\vspace{0.6cm}5\\ \hline 
$\Xi ^{o} $\newline $\Xi ^{-} $ & uss\newline dss & 1314.8\newline 1321.3 & 1297.7 & 4 & 7\\ \hline
$\Sigma *$ & uus,uds,dds & 1385 & 1399.2 & 6 & 9\\ \hline
$\Xi *$ & uss, dss & 1533 & 1532.6 & 10 & 19\\ \hline
$\Omega ^{-}$ & sss & 1672 & 1663.0 & 16 & 31\\ \hline
\end{tabular}
\end{table}
\begin{table}[ht]
\caption{Masses of baryons differing only in the charge q \cite{Griffiths08}}
\begin{tabular}{|p{0.4in}|p{0.5in}|p{2.0in}|p{2.35in}|} \hline 
Baryon & $m_{B}^{\pm } $\newline $MeV/c^2$ & $\Delta m_{B}/m_{B}=(m_{B}^{\pm}-m_{B}^{o})/m_{B}^{o}$ & $\left(\frac{\Delta m_{B} }{m_{B}} \right)/\alpha=\left[(m_{B}^{\pm }-m_{B}^{o})/m_{B}^{o} \right]/\alpha$\\ \hline
\vspace{0.1cm}$N \left\{\begin{array}{l} {p} \\ {n} \end{array}\right.$ &\vspace{0.1cm} 938.272\newline 939.565 &\vspace{0.1cm} $-1.376\cdot 10^{-3} $\newline  &\vspace{0.1cm} -0.188 \\ \hline 
$\Sigma ^{+} $\newline $\Sigma ^{o} $\newline $\Sigma ^{-} $ & 1189.37\newline 1192.64\newline 1197.45 & $-2.74\cdot 10^{-3}$\newline \newline $4.03\cdot 10^{-3} $ & -0.375 \newline \newline 0.552\\ \hline
$\Xi^{o} $& 1314.8 & & \\ $\Xi ^{-}$ & 1321.3 & $4.94\cdot 10^{-3} $ & 0.676\\ \hline 
\end{tabular}
\label{tab:2}
\end{table}
\end{document}